\documentclass[reprint,
superscriptaddress,aps,pre,twocolumn,floatfix,showpacs,amsmath,10pt]{revtex4-1}

\usepackage{booktabs}
\usepackage{xcolor}
\usepackage{graphicx}
\usepackage{graphics}
\usepackage{amsmath}
\usepackage{amssymb}
\usepackage{amsthm,amstext}
\usepackage{amsfonts}
\usepackage{lipsum}
\usepackage{MnSymbol}
\usepackage{mathrsfs}
\usepackage{stmaryrd}
\usepackage{latexsym}
\usepackage[applemac]{inputenc}
\usepackage{accents}

\usepackage[bbgreekl]{mathbbol}
\DeclareSymbolFontAlphabet{\mathbbm}{bbold}
\DeclareSymbolFontAlphabet{\mathbb}{AMSb}
\DeclareMathAlphabet\mathbfcal{OMS}{cmsy}{b}{n}

\newcommand\be{\textbf{\emph{e}}}
\newcommand\bd{\textbf{\emph{d}}}
\newcommand\by{\textbf{\emph{y}}}

\newcommand\bn{\textbf{\emph{n}}}

\newcommand\bi{\textbf{\emph{i}}}

\newcommand\0{\textbf{\emph{0}}}

\newcommand\F{\textbf{F}}

\renewcommand\d\delta
\newcommand\D\Delta
\newcommand\e{\varepsilon}

\newcommand\ph{\varphi}
\renewcommand\r{\varrho}

\newcommand\grad{\text{grad}}

\newcommand\Body{\mathcal{B}}

\newcommand\bbR{\mathbb{R}}

\newcommand\beq{\begin{equation}}
\newcommand\beqn{\begin{eqnarray}}
\newcommand\eeq{\end{equation}}
\newcommand\eeqn{\end{eqnarray}}

\newcommand{\llbra}{\llbracket}
\newcommand{\rrbra}{\rrbracket}

\begin{document}

\title{Stable wrinkling in voltage and charge controlled dielectric membranes}

\author{Giuseppe Zurlo}
\email{giuseppe.zurlo@nuigalway.ie}
\affiliation{Stokes Centre of Applied Mathematics, School of Mathematics, Statistics and Applied Mathematics, NUI Galway, University Road, Galway, Ireland.}

\date{\today}

\begin{abstract}
Thin dielectric elastomers with compliant electrodes exhibit various types of instability under the action of electromechanical loading. Guided by the thermodynamically-based formulation of Fosdick and Tang (J. Elasticity 88, 255-297, 2007), here we provide an energetic perspective on the stability of dielectric elastomers and we highlight the fundamental energetic divide between voltage control and charge control. By using the concept of energy relaxation, we describe wrinkling for neo-Hookean ideal elastomers, and we show that in voltage control wrinkling is stable as long as the tension-extension inequality holds, whereas wrinkling is always stable in charge control. We finally illustrate some examples involving both homogeneous and inhomogeneous deformations, showing that the type and hierarchy of instabilities taking place in dielectric membranes can be tuned by suitable choices of the boundary conditions. 
\end{abstract}

\maketitle

\section{Introduction}

The electromechanical behavior of thin dielectric elastomers, together with the type and order of instabilities affecting their performances, is fundamentally different in voltage and in charge controlled systems. Shedding light onto such divide is important for the design of dielectric elastomers in applications like sensing, actuation, energy harvesting and on-demand patterning \cite{Carpi,CSZ14,CSZ15}. 

Under the action of slowly increasing electric fields, deformations of dielectric elastomer membranes grow up, until a catastrophic thinning (known as electromechanical or {\it pull-in} instability) {\it may} take place -- an outcome that preludes to the device failure \cite{PRL17}.  Electromechanical instability is deceivingly related to wrinkling, which is due to the relaxation of in-plane compression through out-of-plane finely oscillating deformations. Wrinkling, per se, can be {\it stable}, in the sense that it does not necessarily lead to failure. This is an important feature, since it discloses the technological possibility to exploit on-demand patterns by maintainable wrinkled states \cite{Pattern1,Pattern2,Kollosche, Liu, Mao18a, Mao17, Mao18b}. Such behaviors are, however, very different in voltage and charge controlled systems. 

The purpose of this article is to provide a physical insight into the stability of dielectric elastomers under both voltage and charge controls, and to elucidate the subtle connection between wrinkling and pull-in instabilities in both cases (see also \cite{DPSZ,Greaney18}, which refer to specific geometries). Guided by the thermodynamically-based energetic formulation of Fosdick and Tang \cite{FT}, we assume that in close-to-equilibrium conditions stable configurations should minimize a suitable electromechanical energy, and we highlight the differences between voltage and charge controls. After showing that wrinkling can be consistently described by the rank-one relaxation of the electroelastic potential, we then show that the requirement of electromechanical stability implies the existence of both stable and unstable wrinkling regimes. When our analysis is specialised to ideal dielectrics with neo-Hookean elastic response, we find that wrinkling may loose stability in voltage control, whereas wrinkling is always stable in charge control. 

The idea of energy relaxation can be used to solve general boundary value problems, and we conclude this article by proposing some possible experiments involving both homogeneous and inhomogeneous deformations \cite{Mao18a,Greaney18}, which show that a judicious choice of boundary conditions can provide stabilization of wrinkled patterns.

\section{Electroelastic stability}

\subsection{Energy minimization}

\begin{figure}[htb]
\begin{centering}
\includegraphics[width=5.5cm]{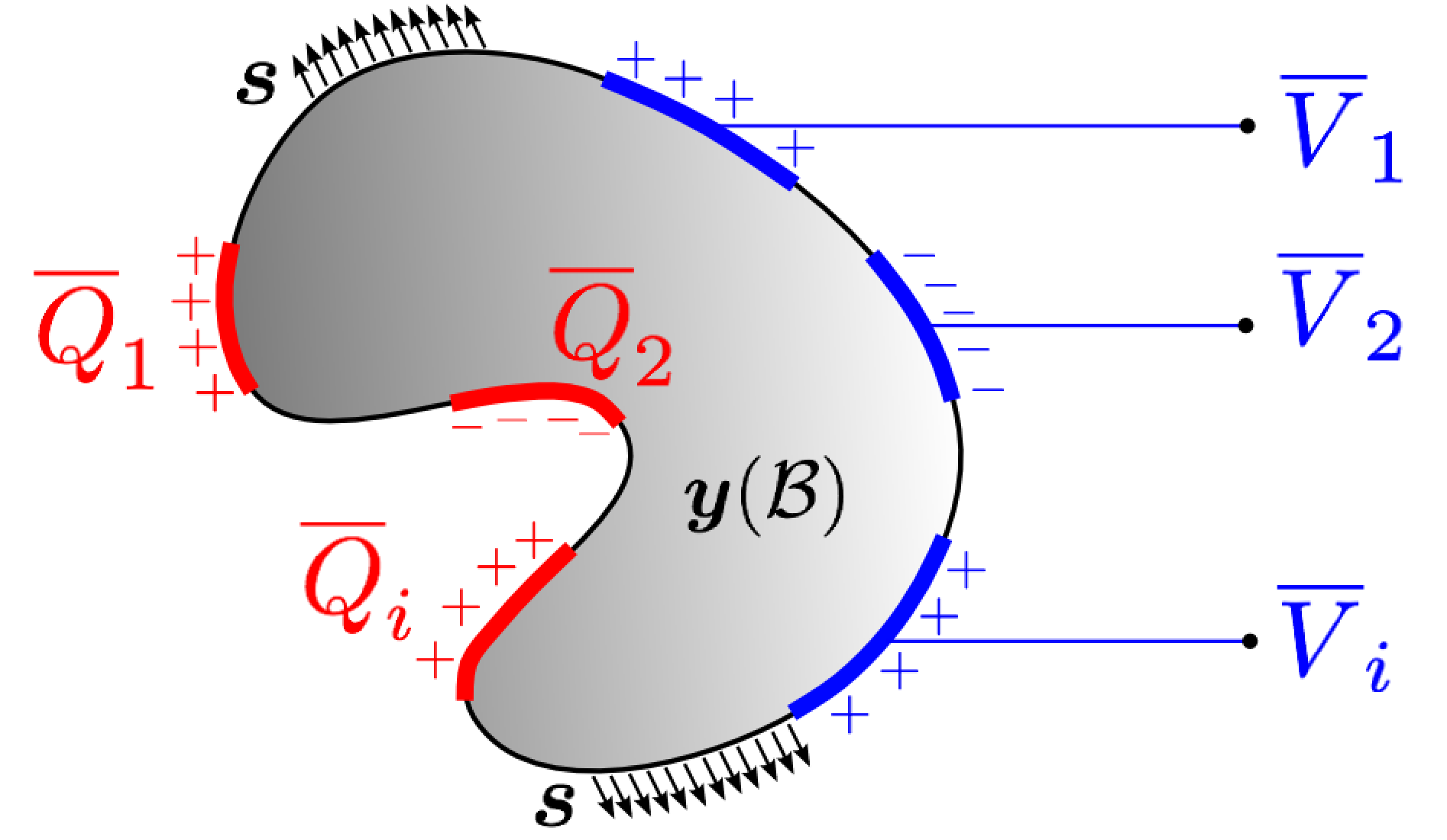}
\par\end{centering}
\caption{\label{scheme} A schematic representation of the current configuration $\by(\Body)$ of a dielectric elastomer under the combined action of tractions $\boldsymbol{s}$ and electrostatic controls. More in detail, the total charges $\overline{Q}_i$ are controlled on $n_q$ conductors and the voltages $\overline{V}_i$ are controlled on $n_v$ conductors. In both cases, the compliant conductors are glued on the body boundary. }
\end{figure}

Take into consideration a system of $n=v+q$ compliant conductor surfaces $\mathcal{C}_i$, $v$ of them where voltages $\overline{V}_i$ are controlled, while their total charge $Q_i$ is left free; and $q$ of them where the total charges $\overline{Q}_i$ are controlled, while their voltage $V_i$ is left free. Full voltage control corresponds to $q=0$ whereas full charge control corresponds to $v=0$. Consider also a stress-free configuration $\Body$ for the elastic dielectric and assume that all the conductors are compliant sub-parts of the boundary surface $\partial\Body$ of the dielectric, see Fig.\ref{scheme}. The conductors shapes, henceforth, are fully determined by the body deformation. Under these assumptions, an electroelastic state defined by the couple $(\by,\bd)$, where $\by$ is the deformation and $\bd$ the (Eulerian) electric displacement, is stable if it minimizes the electromechanical energy \cite{FT}
\beq\label{energy0}
\mathcal{E} =
\int_{\Body}\psi(\nabla\by,\bd)\,dv + 
\int_{\text{vac}}\omega(\bd)\,dv - 
\sum_{i=1}^{v} Q_i \overline{V}_i - \mathcal{W}^m
\eeq
where $\nabla\by$ is the deformation gradient, $\psi$ is the electroelastic energy of the deformable dielectric, $\omega=\be_0\cdot\bd_0/2$ is the electric energy of the vacuum $\text{vac}\equiv\bbR^3\backslash\Body$ surrounding the body, where $\be_0$ is the electric field in the vacuum, where $\bd_0=\e_0\be_0$ and where $\epsilon_0$ is the vacuum permittivity. Finally, $\mathcal{W}^m$ is the mechanical work. 

We further focus on incompressible \emph{linear ideal dielectrics}, where elastic and electric effects are constitutively uncoupled \cite{SuoGreene08,ZDL}. Here $\bd=\e\be$, where $\e$ is the dielectric permittivity and $\be=-\grad\ph$ is the electric field, with $\ph$ the electrostatic potential. In this case,  the electroelastic energy additively splits as $\psi=w(\nabla\by)+\be\cdot\bd/2$, where $w$ measures the purely elastic response of the elastomer. By using integration by parts, together with the fact that $Q_i=\int_{\by(\mathcal{C}_i)}\sigma_i\,da$, where the charge density on each conductor is $\sigma_i=\llbra\bd_i\rrbra\cdot\bn$ with $\bn$ the outward normal to $\by(\partial\Body)$, and by imposing that $\text{div}\bd=\0$ and that the electrostatic potential at infinity decays to zero, the electromechanical energy \eqref{energy0} may be recast as
\beq\label{energy1}
\mathcal{E}  =
\int_{\Body}w(\nabla\by)\,dV + \frac{1}{2}
\sum_{i=1}^{q} \overline{Q}_i V_i - 
\frac{1}{2}\sum_{i=1}^{v} Q_i \overline{V}_i - \mathcal{W}^m. 
\eeq
This expression highlights a first important divide between voltage control and charge control. Neglecting for a moment elastic effects, when voltage is controlled $(q=0)$ energy minimization requires the sum $\tfrac{1}{2}\sum Q_i \overline{V}_i$ to be {\it maximized}. When charge is controlled $(v=0)$, then energy minimization requires the sum $\tfrac{1}{2}\sum\overline{Q}_i V_i$ to be {\it minimized}, which implies that free charges are arranged on the electrodes in such a way as to minimize the total electrostatic energy; this is consistent with the classical formulation of the Thomson's theorem of electrostatics \cite{LL,Kovetz}. This divide between voltage and charge controls is also consistent with the subtle transition between energy based minimization and enthalpy based saddle principles in electroelasticity \cite{Miehe11,DO,DO19}. 

\subsection{Stability in voltage and charge controls}

From now on our analysis will be confined to homogeneous deformations. We assume that the reference, stress-free configuration $\Body$ is a thin prismatic region with mid-surface $\Omega$ of area $A$ and uniform thickness $H$. Two fully compliant electrodes are glued on the upper and lower faces of the dielectric. We assume that the gradient of deformation can be written as $\nabla\by=\sum_{i=1}^3\lambda_i\bi_1\otimes\bi_i$, where $\lambda_i$ with $(i=1,2,3)$ are the principal stretches along the mutually perpendicular directions $\bi_i$. where $\lambda_{1/2}$ are the stretches in the plane of the membrane, whereas due to incompressibility the thickness stretch is $\lambda_3=1/(\lambda_1\lambda_2)$. We further denote by $(s_1,s_2)$ the principal components of the in-plane stress Piola-Kirchhoff stress. The total potential energy functional \eqref{energy1} thus specializes as $\mathcal{E}=\psi-s_1\lambda_1-s_2\lambda_2$, where (by neglecting inessential constants) the electroelastic potential in voltage control is
\beq\label{EE}
\psi(\lambda_1,\lambda_2;E) = w(\lambda_1,\lambda_2) - \frac{E^2}{2} (\lambda_1\lambda_2)^2 
\eeq
where $E=V/H$ is the only non-vanishing component of the Lagrangian electric field in the thickness direction. When the total charge $Q$ is controlled instead, the electroelastic potential reads
\beq\label{ED}
\psi(\lambda_1,\lambda_2;D) = w(\lambda_1,\lambda_2) + \frac{D^2}{2(\lambda_1\lambda_2)^2} 
\eeq
where $D=Q/A$ is the Lagrangian electric displacement in the thickness direction. In both cases we have set $\e=1$ for simplicity. 

In this simplified setting, the minimization of \eqref{energy1} specialises as follows: for a prescribed triple $(s_1,s_2,E)$ in voltage control, or for a prescribed triple $(s_1,s_2,D)$ in charge control, find the stationary stretches $(\lambda_1^{\text{e}},\lambda_2^{\text{e}})$ such that 
\beq\label{stress}
s_{\alpha} = \left[\frac{\partial\psi(\lambda_1,\lambda_2) }{\partial\lambda_{\alpha}}\right]_{(\lambda_1^{\text{e}},\lambda_2^{\text{e}})}\qquad (\alpha=1,2)
\eeq
and for which $\psi(\lambda_1,\lambda_2)$ is locally convex \cite{Ericksen}. Albeit it is well known that convexity is an overly restrictive condition in the description of large deformations \cite{Ball}, the incipient lack of convexity at a stationary homogeneous state is identified with the so-called {\it pull-in} or {\it electromechanical instability} in the literature on dielectric elastomers \cite{SuoSinica}. A more refined discussion on stability requires to account for inhomogeneous deformations \cite{DO,SharmaPull2010}, but since these are out of the scope of our study, we here identify electromechanical instability with lack of convexity of the electroelastic energy at stationary states. 
\begin{figure}[htb]
\begin{centering}
\includegraphics[width=\columnwidth]{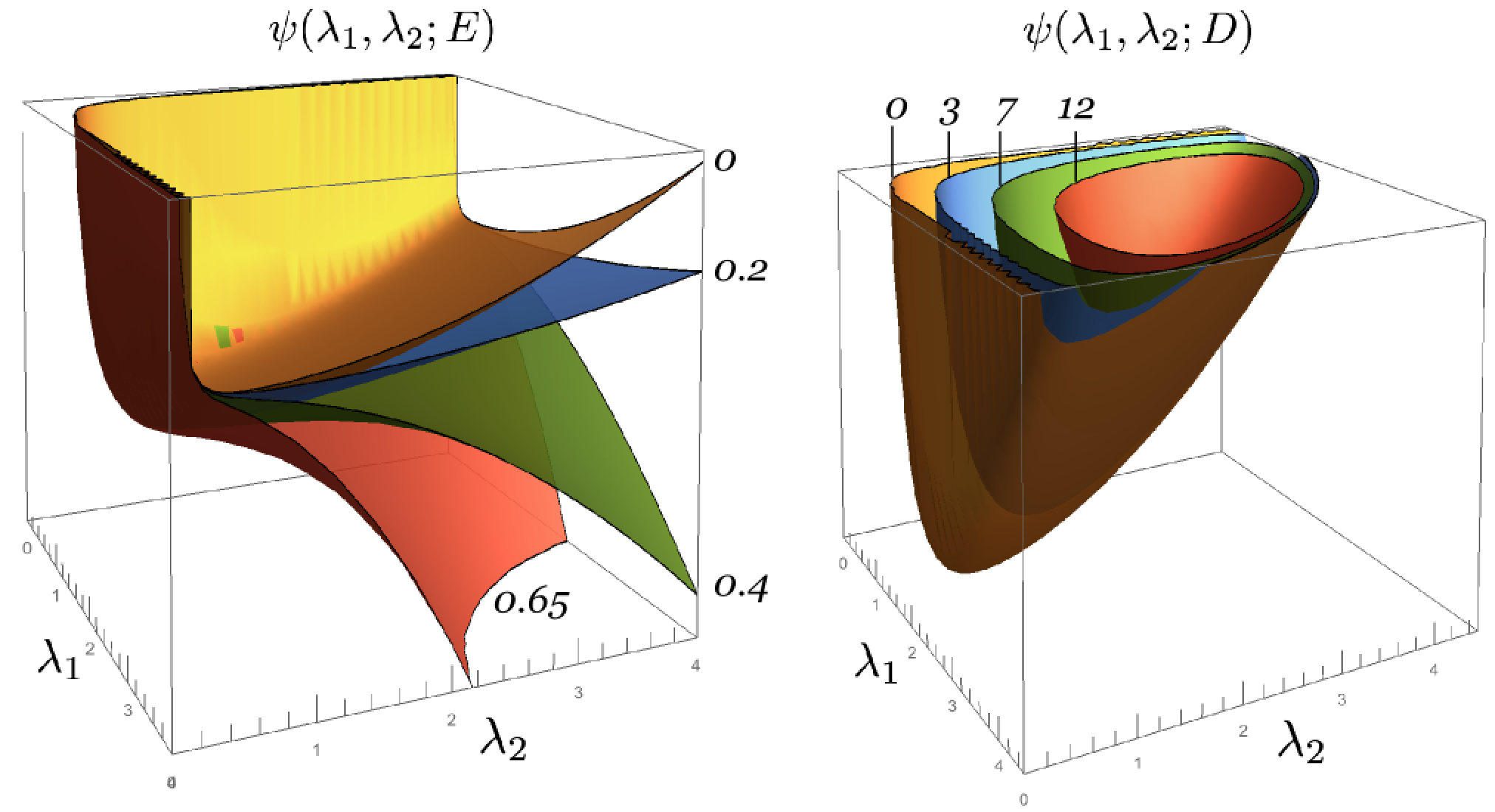}
\par\end{centering}
\caption{\label{energyfig} Left, electroelastic energies in voltage control for various values of $E$ and, at the right,  in charge control for various values of $D$. The elastic part $w=(\lambda_1^2+\lambda_2^2+\lambda_1^{-2}\lambda_2^{-2}-3)/2$ is neo-Hookean.}
\end{figure}

To further focus only on electrically induced effects we discard the possibility of purely mechanical instabilities, and from now on we assume that $w$ is convex. With this assumption, inspection of \eqref{EE},\eqref{ED} succinctly explains the major divide between voltage control and charge control in terms of stability. 
\begin{figure}[htb]
\begin{centering}
\includegraphics[width=4.5cm]{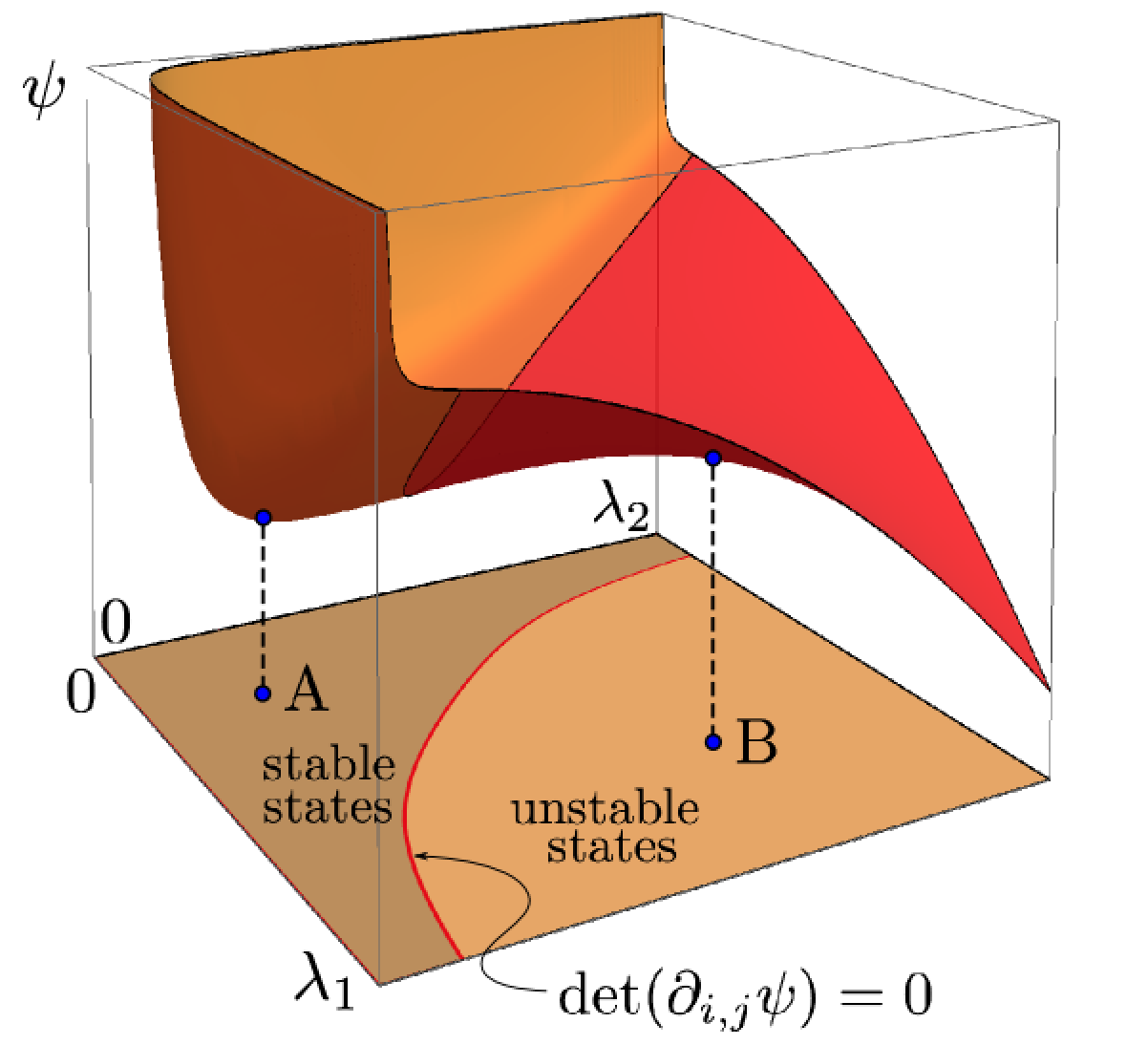}
\par\end{centering}
\caption{\label{Econv} In voltage control and for a neo-Hookean energy, stable states are those where the electroelastic energy is convex. This partitions the plane $(\lambda_1,\lambda_2)$ into stable (darker) and unstable (lighter) regions, according to \eqref{Ecrit}. Points $A$ and $B$, where $s_{\alpha}=0$, correspond to stable and unstable tensionless states, respectively. Here $E=0.3$.}
\end{figure}

In voltage control, equation \eqref{EE} reveals that the term $-(E\lambda_1\lambda_2)^2/2$, which is non-convex and unbounded from below, introduces a source of non-convexity in the total potential energy, see Fig.\ref{energyfig}. Thus, a stationary solution of \eqref{stress} may become unstable as soon as the Hessian $\partial_{i,j}\psi$ ceases being positive definite, where $\partial_i$ denotes partial differentiation with respect to $\lambda_i$. This is consistent with the so-called {\it Hessian criterion} \cite{SuoGreene08,SuoSinica,FuDorfmann,ZhaoSuoHessian}. For example, for neo-Hookean materials one finds that for a given $E$, stable states of stretch are those in the stable region (see also Fig.\ref{Econv}) 
\beq\label{Ecrit}
3E^4 + 
E^2\frac{\lambda_1^2\lambda_2^4 + \lambda_1^4\lambda_2^2 -2}{\lambda_1^4 \lambda_2^4}
-\frac{5+3\lambda_1^2\lambda_2^2 (\lambda_1^2+\lambda_2^2)}{\lambda_1^8\lambda_2^8}\leq1. 
\eeq
Relative to charge control, we see from \eqref{ED} that the since $(D/(\lambda_1\lambda_2))^2/2$ is a strictly convex function of the principal stretches, and since the sum of convex functions gives another convex function, the total electroelastic energy $\psi$ remains a convex function of the principal stretches for all values of $D$, see Fig.\ref{energyfig}. Since convexity can not be lost in this case, electromechanical instability does not exist in charge control - clearly, as long as one assumes that the purely elastic energy $w$ is convex.

From now on, our analysis will be specialised to neo-Hookean materials.

\subsection{The stability of tensionless states\label{SecTensless}}

A special connection exists between electromechanical instability and tensionless states, that are characterised by $s_1=s_2=0$. With $w=(\lambda_1^2+\lambda_2^2+\lambda_1^{-2}\lambda_2^{-2}-3)/2$, one easily finds that in voltage control, if $E\neq 0$, there are two stress-free states $A=(\lambda_0,\lambda_0)$ and $B=(\lambda'_0,\lambda'_0)$, with $\lambda_0\leq\lambda_0'$, where $A$ corresponds to a local minimum of $\psi$ (thus it is stable), whereas $B$ to a saddle (thus it is unstable), see Fig.\ref{Econv}. In particular there is a limit value $E^{\text{lim}}=\sqrt{3}/2^{4/3}\simeq 0.687$ for which $A\equiv B$. Above such limit, the electroelastic energy possesses no local minimizers. 

Oppositely, in charge control the electroelastic energy remains convex for all values of $D$, henceforth there is only one point  where $s_1=s_2=0$ and such point is always stable. Conclusively, tensionless states are always stable in charge control, whereas they can lose stability in voltage control. In voltage control, the connection between tensionless states and pull-in was already provided in \cite{APL2011,IJNLM2012}, and it was recently confirmed in \cite{Greaney18,Su2018}.


\section{Electrically induced wrinkles}

\subsection{Tension field theory for dielectric elastomers}

The discussion above is incomplete when dealing with thin elastomers, that wrinkle immediately at the outset of membranal compression. Based on the partition of the plane $(\lambda_1,\lambda_2)$ into stable and unstable regions as done above, one could attempt partitioning the same plane into a region $\mathcal{S}$ of {\it taut states} (where both principal stresses are positive), regions $\mathcal{U}_{\alpha}$ ($\alpha=1,2$) of {\it wrinkled states} (where only one principal stress is negative) and a {\it tensionless states} $\mathcal{L}$ where both principal stresses are negative \cite{Pipkin86fabrics}. Similar partitions were given in \cite{APL2011,IJNLM2012} for dielectric membranes and in \cite{DeSimone,Cesana} for  nematic elastomers.

Such partition, however, requires to modify the original energy $\psi$ in order to account for wrinkling. This target is elegantly accomplished through {\it tension field theory} \cite{Steigmann,Mansfield,Pipkin86,Pipkin93}, that embeds the internal constraint of lack of resistance to compression by replacing the parent energy $\psi$ with a {\it relaxed energy} $\psi^*$. In the relaxed energy, a stress component is automatically set to zero  whenever it would be negative in the parent energy \cite{Pipkin86,Steigmann}. 

To describe the construction of the relaxed electroelastic energy, we follow closely the argument of \cite{Pipkin86}. After recalling that $w$ of a neo-Hookean material is strictly convex, consider first the region of stretches $\lambda_1\geq\lambda_2\geq\lambda_0$ and assume that a value of $\bar\lambda_1$ is prescribed such that $s_1(\bar\lambda_1,\lambda_2)>0$. Suppose that $\lambda^*_2(\bar\lambda_1)$ is the stretch where $\psi(\bar\lambda_1,\lambda_2)$ is minimum, that is where $s_2(\bar\lambda_1,\lambda_2^*(\bar\lambda_1))=0$. The stretch $\lambda_2^*(\bar\lambda_1)$ is the {\it natural width in simple tension} \cite{Pipkin86}, and $s_2$ is positive or negative if $\lambda_2-\lambda_2^*(\bar\lambda_1)$ is positive or negative, respectively. In voltage and charge control, one easily finds that, respectively, 
\beq\label{lambdanat}
\lambda_2^*(\lambda_1) = \frac{1}{\left(\lambda_1^2 - E^2\lambda_1^4\right)^{1/4}}, \quad
\lambda_2^*(\lambda_1) = \frac{\left(1+D^2\right)^{1/4}}{\sqrt{\lambda_1}}. 
\eeq
\begin{figure}[htb]
\begin{centering}
\includegraphics[width=\columnwidth]{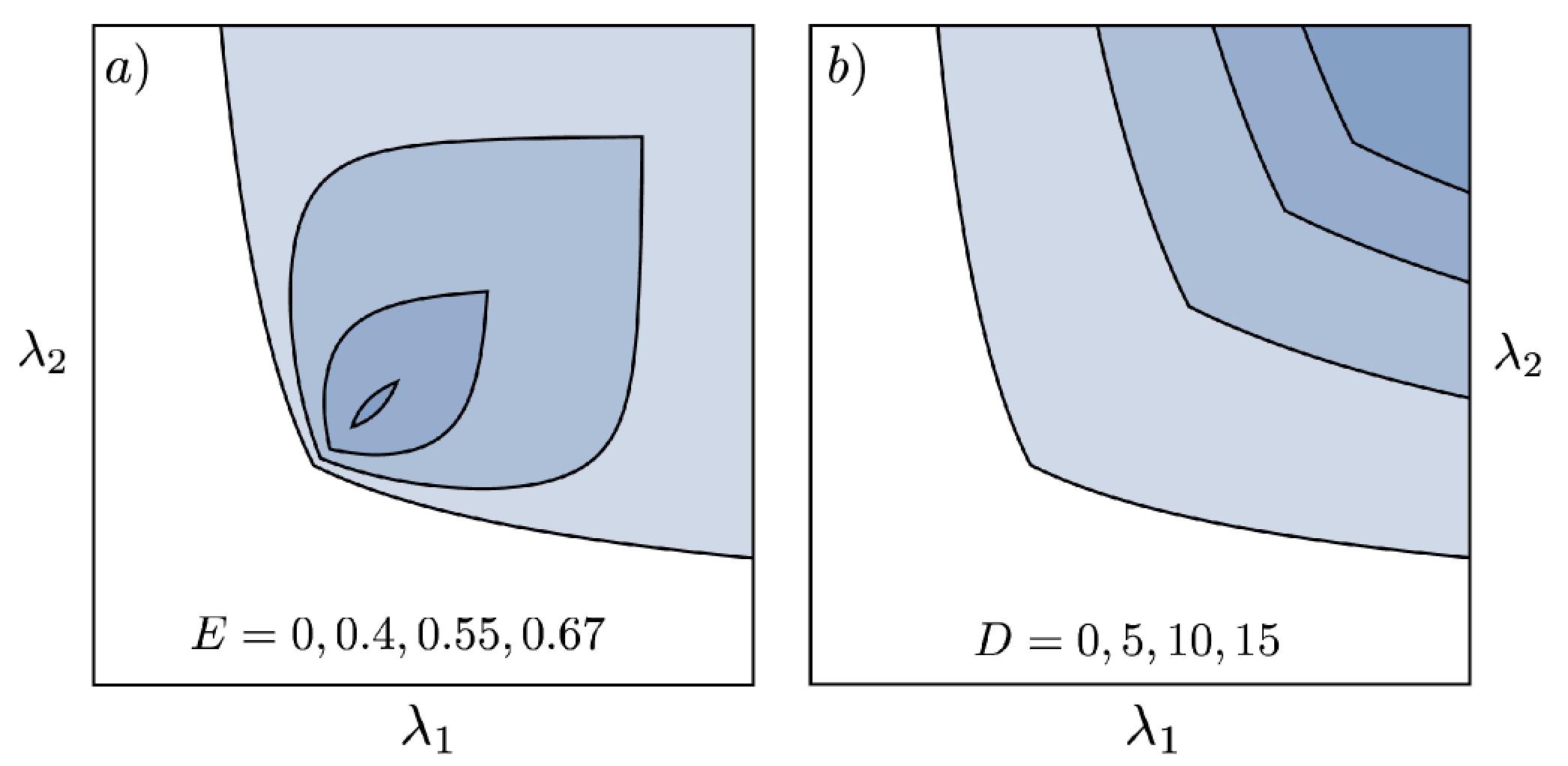}
\par\end{centering}
\caption{\label{naturalwidths} Evolution of the region $\mathcal{S}$ of taut states in the $(\lambda_1,\lambda_2)$ plane in voltage $a)$ and in charge $b)$ control. Taut regions are denoted by shaded areas and darker colors refer to higher values of the controlled $E$ or $D$. In voltage control the region of taut states shrinks down and it completely disappears for $E>E^{\text{lim}}=0.687$. The taut region is simply shifted towards higher stretches in charge control.}
\end{figure}
Since the membrane is thin, it wrinkles as soon as $s_2<0$, that means, as soon as $\lambda_2<\lambda_2^*(\bar\lambda_1)$. Tension field theory dictates that whenever the membrane is wrinkled in a certain direction, further shortening along the same direction does not alter its energy. This implies that in the region $\mathcal{U}_1$ where $\lambda_1\geq\lambda_2\geq\lambda_0$ and $\lambda_2\leq\lambda_2^*(\bar\lambda_1)$, the electroelastic energy remains ``frozen'' at its value at the outset of wrinkling, given by $\psi^*(\lambda_1)\equiv\psi(\bar\lambda_1,\lambda_2^*(\bar\lambda_1))$. In this region the stress now calculates as $s^*_{\alpha}=\partial\psi^*/\partial\lambda_{\alpha}$, so that $s_1^*$ is only a function of $\lambda_1$ and $s_2^*=0$, as desired. Clearly, the same argument applies in the region $\mathcal{U}_2$ where $\lambda_2\geq\lambda_1\geq\lambda_0$, whereas no modification of the parent energy is required in the region $\mathcal{S}$ where both the principal stresses computed from $\psi$ are positive. 

Finally, the membrane is completely tensionless in the region $\mathcal{L}$ where both principal stretches are lower than $\lambda_0$, or (only in voltage control) in the region $\mathcal{L}'$ where both principal stretches are higher than $\lambda'_0$. In such regions the relaxed energy should amount to a constant, but special care is required when dealing with the region $\mathcal{L}'$. Indeed, the relaxed energy $\psi^*$ must be the largest increasing function of $(\lambda_1,\lambda_2)$ that is lower than $\psi$, see \cite{PipkinARMA} for details. This implies that the construction of the electroelastic energy should actually be confined to the regions $(\mathcal{L},\mathcal{S},\mathcal{U}_{\alpha})$, since for $\lambda_{\alpha}>\lambda_0'$ the energy $\psi$ is a decreasing function. Conclusively, the relaxed electroelastic energy can be constructed as
\beq\label{relaxed}
\psi^*(\lambda_1,\lambda_2)= 
\left\{
\begin{array}{rccccc}
\psi(\lambda_0,\lambda_0) & \text{in} & \mathcal{L}\\
\psi(\lambda_1,\lambda_2^*(\lambda_1)) & \text{in} & \mathcal{U}_1\\
\psi(\lambda_1^*(\lambda_2),\lambda_2) & \text{in} & \mathcal{U}_2\\
\psi(\lambda_1,\lambda_2) & \text{in} & \mathcal{S}\\
\text{undefined} & \text{in} & \mathcal{L}'
\end{array}
\right.
\eeq
where the last line only applies in voltage control. 

The difference between voltage and charge control in terms of wrinkling is due to the peculiar evolution of the taut region $\mathcal{S}$ with increasing $E$ and $D$, see Fig.\ref{naturalwidths}. Indeed, Eq.\eqref{lambdanat} shows that in charge control $\lambda_2^*$ is a decreasing function of $\lambda_1$, whereas in voltage control such function depends non-monotonically on $\lambda_1$. This is consistent with the fact that in charge control there is only one tensionless state defined by $\lambda_2^*(\lambda_1)=\lambda_1$, whereas there are two such states in voltage control (see Sec.\ref{SecTensless}). 

The different behavior of $\lambda_2^*$ in voltage and charge control implies that in the former case the region of taut states $\mathcal{S}$ is closed, with apexes defined by $A$ and $B$, whereas such region is always open in charge control, see Fig.\ref{naturalwidths}. Since in voltage control the points $A$ and $B$ converge for growing $E$ and finally disappear for $E>E^{\text{lim}}$, there exists no stable state above such limit electric field, see \cite{APL2011,IJNLM2012} for this analysis with other types of constitutive relations. 

Such behavior is completely absent in charge control, where the action of electric fields simply shifts the region of taut states towards higher stretches. This is consistent with the analysis on the stability of tensionless states done in Sec.\ref{SecTensless}, and with the findings of \cite{Michelsubmitted}.

\subsection{Stable wrinkling in voltage and charge controls}

In the perspective of energy minimization, also in the presence of wrinkling a stationary state is stable if the relaxed energy is convex. In charge control, at least for neo-Hookean materials, one immediately verifies that the relaxed energy is convex for all values of the stretch, meaning that all taut, wrinkled and tensionless states are always stable. In voltage control, instead, the convexity requirement leads to further partitioning of the region of taut, wrinkled and tensionless states into stable and unstable regions.

\begin{figure}[htb]
\begin{centering}
\includegraphics[width=\columnwidth]{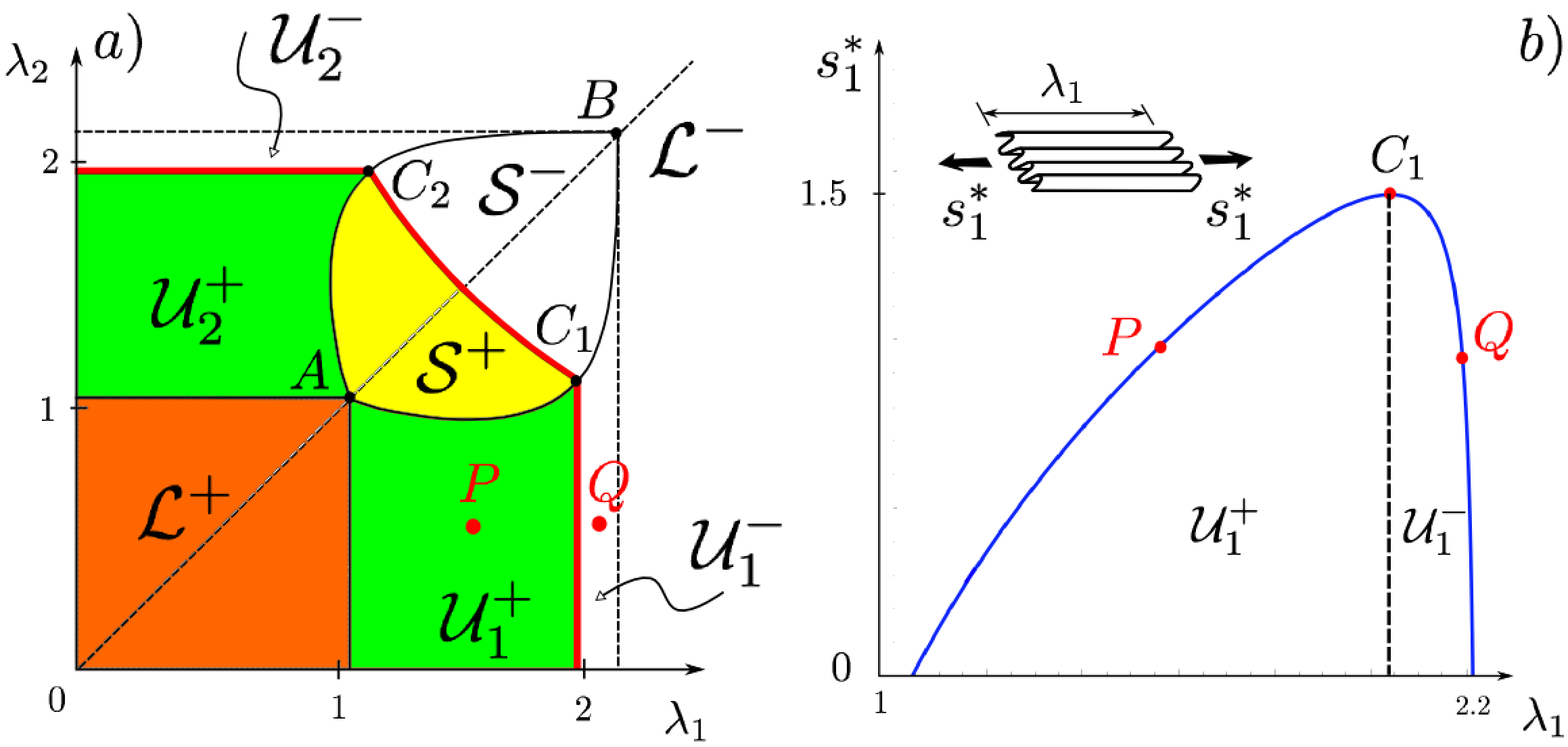}
\par\end{centering}
\caption{\label{softening} $a)$ partition of the $(\lambda_1,\lambda_2)$ plane into stable $(+)$ and unstable $(-)$ regions for $E=0.45$. The red curve corresponds to the incipient lack of convexity of the relaxed energy. $b)$ stress-stretch response in the wrinkled region $\mathcal{U}_1$ into a stable region $\mathcal{U}_1^+$ where the tension-extension inequality is satisfied, and into an unstable region $\mathcal{U}_1^-$ where this condition is not satisfied. Points $P$ and $Q$ are stable and unstable wrinkled states, respectively.}
\end{figure}

In the region $\mathcal{S}$ of taut states it results that $\psi^*\equiv\psi$, so stability reduces to the inspection of the parent energy. In this case the requirement of convexity partitions $\mathcal{S}$ into a stable region $\mathcal{S}^+$ (containing $A$) where the Hessian $\partial_i\partial_j\psi$ is positive definite, and into an unstable region $\mathcal{S}^-$ (containing $B$) where the Hessian is not positive definite, see also Fig.\ref{Econv}. The boundary curve between $\mathcal{S}^+$ and $\mathcal{S}^-$, where $\det\partial_i\partial_j\psi=0$, intersects the edges $\lambda_2=\lambda_2^*(\lambda_1)$ and $\lambda_1=\lambda_1^*(\lambda_2)$ in two points, that we denote by $C_1$ and $C_2$, respectively, see Fig.\ref{softening}$_a$. 

In the region $\mathcal{U}_1$ the relaxed energy depends only on $\lambda_1$, so the requirement of convexity of $\psi^*$ reduces to the requirement that $\partial_1 s_1^*\geq 0$. Such condition, imposing that $s_1^*$ must be an increasing function of $\lambda_1$, is nothing but the {\it tension-extension} inequality \cite{TN} along the non-wrinkled direction $\bi_1$. Note that since the value of $\lambda_1$ where $\partial_1 s_1^*= 0$ coincides precisely with $\lambda_1(C_1)$, the region $\mathcal{U}_1$ can be partitioned into a region $\mathcal{U}_1^+$ of {\it stable wrinkling}, where  $\lambda_1\in(\lambda_0,\lambda_1(C_1))$, and into the region $\mathcal{U}_1^-$ of {\it unstable wrinkling}. In other words, the manifestation of electromechanical instability in the wrinkled region takes place through mechanical softening, see Fig.\ref{softening}$_b$. An analogous partition holds for $\mathcal{U}_2$. 

The stability of the tensionless region $\mathcal{L}$ was already analysed in Sec.\ref{SecTensless}. Such region is always stable as long as the electric field is lower than the limit electric field $E^{\text{lim}}$ where $A\equiv B$. For what concerns the region $\mathcal{L}'$, even though the relaxed energy is not properly defined in this state, the energy of any point in this region is equal to the energy of the point $B$ and, henceforth, it is unstable. For this reason, we thus denote $\mathcal{L}$ by $\mathcal{L}^+$ and $\mathcal{L}'$ by $\mathcal{L}^-$. 

\section{Possible experiments}

\subsection{Homogeneous deformations}

\begin{figure}[htb]
\begin{centering}
\includegraphics[width=\columnwidth]{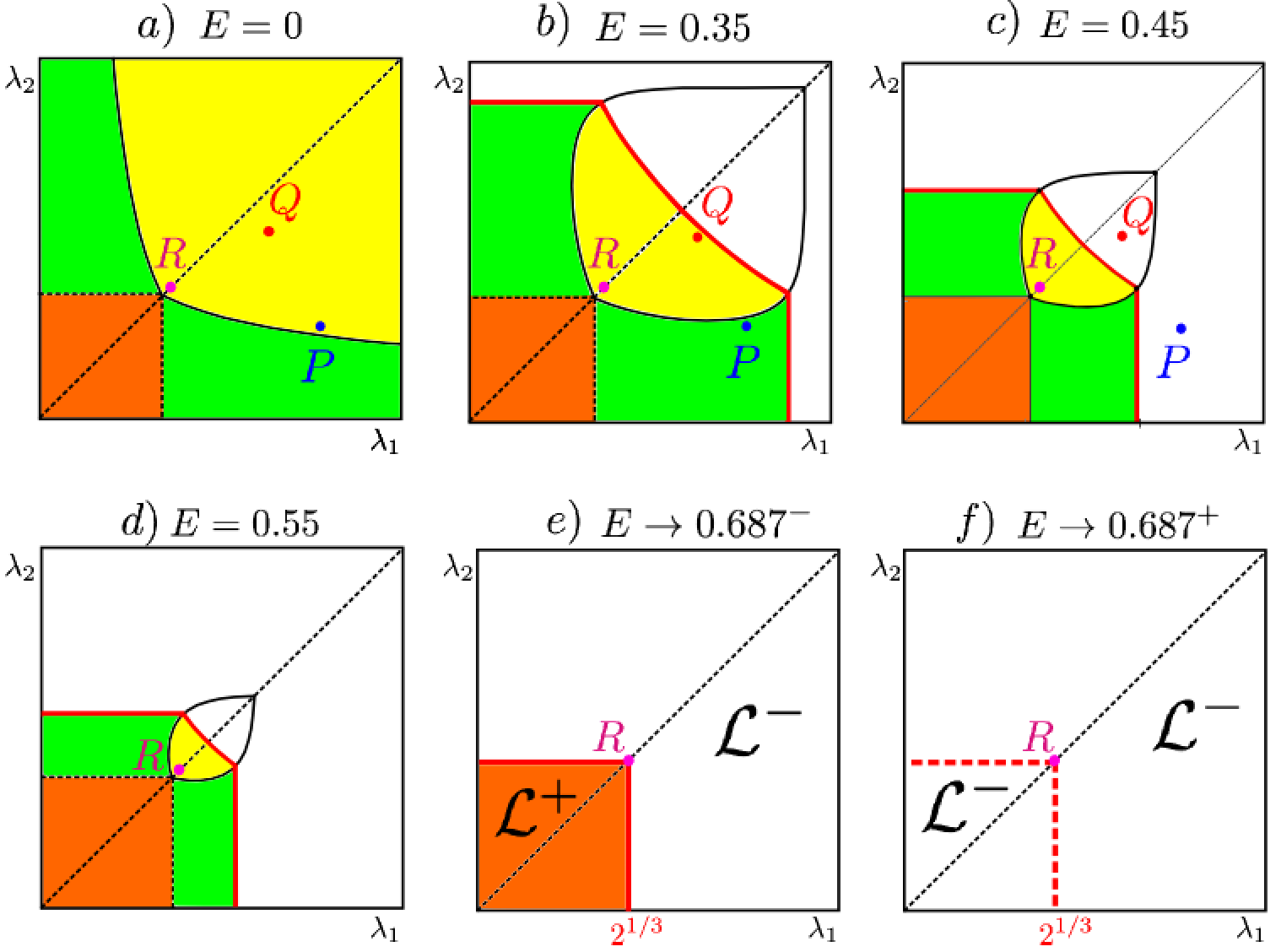}
\par\end{centering}
\caption{\label{experiment} Evolution of the stable taut and wrinkled regions in voltage control.}
\end{figure}

By focussing on voltage control, we now provide three experiments involving homogeneous deformations that can be used to assess the predictive power of the ideas discussed in this article. Assume that when $E=0$ a unit square of dielectric membrane is prestretched, by imposing edge displacements, into a rectangular membrane with edges of length $(\lambda_1,\lambda_2)$. To represent three different scenarios, the level of prestretch is described by the points $P,Q,R$ in Fig.\ref{experiment}. 

Consider first the point $P$. When $E=0$ (inset $a$) the membrane is taut in both directions and stable, since $P$ falls in the region $\mathcal{S}^+$. When voltage is further increased (inset $b$), the membrane now falls into the region $\mathcal{U}_1^+$, so it wrinkles along $\bi_1$, and such homogenous configuration is stable. Finally (inset $c$), the rectangular membrane will undergo electromechanical instability {\it while being wrinkled}, when by further increasing the electric field the point $P$ falls in the region $\mathcal{U_1^-}$. This is an example that highlights the possibility that pull-in could be reached after that wrinkling has occurred. 

Consider now the state described by the point $Q$. Initially (inset $a$) the membrane is taut in both directions. Then, as voltage is increased, electromechanical instability is reached while the membrane still taut, so no wrinkling has yet taken place. 

Consider finally point $R$, that is prestretched at a level $(2^{1/3},2^{1/3})$. In this case the membrane remains taut and stable up until the electric field reaches the critical value $E=0.687$. Here, the membrane undergoes at the same time to electromechanical instability and conformal loss of tension - meaning that the membrane will exhibit a random wrinkling pattern in all directions. 

\subsection{Inhomogeneous deformations}

We now apply the theory to study inhomogeneous deformations and the possible coexistence of wrinkled and non-wrinkled regions. In such conditions, by finely tuning boundary conditions and the electric field, stability can be lost either through loss of convexity in the taut part of the membrane while the wrinkled part is still stable, or vice-versa. 

Consider a dielectric membrane with compliant electrodes that, in its stress-free configuration, is shaped as a flat disk with internal and external radii $(r_i,r_e)$, respectively. The whole upper and lower surfaces of the disk are electrically controlled by the application of a voltage. Confine attention to axially symmetric deformations and denote by $\r(r)$ the radial deformation, so that the principal stretches in the radial and hoop directions are $\lambda_r=\r'$ and $\lambda_{\theta}=\r/r$, respectively. The radial and hoop components $s_{r/\theta}$ of the Piola-Kirchhoff stress tensor can be calculated as derivatives of the electroelastic energy both in the taut region and in the wrinkled region. 

\begin{figure}[htb]
\begin{centering}
\includegraphics[width=7.5cm]{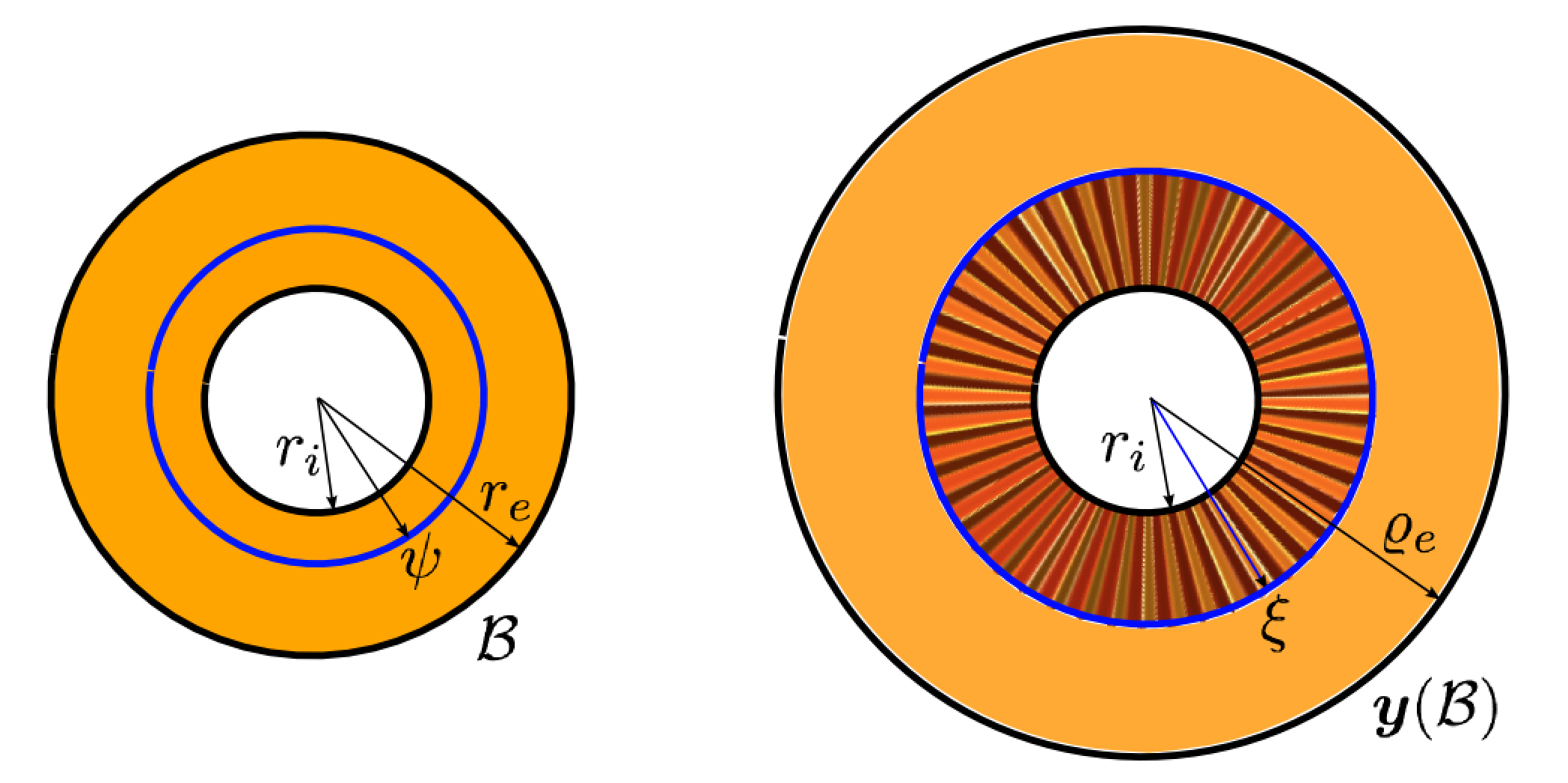}
\par\end{centering}
\caption{\label{disk} Schematic representations of the stress-free configuration $\Body$ (left) and current configuration $\by(\Body)$ (right) of a thin disk actuated through the application of an electric field on the whole upper and lower surfaces. The radii $\psi$ and $\xi$ correspond to 
the referential and current placements, respectively, between wrinkled (internal) and non-wrinkled (external) regions.}
\end{figure}

Assume that, prior to the application of a voltage, the external rim is displaced to $\r(r_e)=\r_e>r_e$ and kept fixed in this position, whereas the internal rim is kept fixed at $\r(r_i)=r_i$, see Fig.\ref{disk}. 

For a neo-Hookean membrane, as long as $E=0$ both stress components are inhomogeneous and positive, so the membrane is not wrinkled. As an electric field is applied, the membrane undergoes an in-plane expansion, that results into stress relaxation. In particular, since $s_{\theta}<s_{r}$ and since $s_{\theta}$ is lower towards $r_i$, the internal part of the disk will wrinkle first - clearly, if stability is not lost even before developing wrinkles. Then, as the electric field is further increased, the referential phase-boundary located at $r=\psi$ will propagate from $r_i$ towards $r_e$ (or, in the current configuration, the current phase boundary located at $\r=\xi$ will propagate from $r_i$ towards $\r_e$).  

In the absence of wrinkling the problem is trivial. In the presence of wrinkling, the equilibrium problem consists in solving a coupled system of second-order ordinary differential equations for $(\r,\r^*)$. Equilibrium in the non-wrinkled region $(\psi,r_e)$ dictates that $s_{r}'+(s_{r}-s_{\theta})/r = 0$, that provides a non-linear second order differential equation for $\r$. In the wrinkled region it results $s^*_{\theta}=0$ whereas $s^*_r$ is a function of $\lambda_r^*={\r^*}'$ only, where $\r^*$ is the radial deformation in $(r_i,\psi)$. In this region equilibrium reduces to $(r s^*_r)'=0$. Since both $\psi$ and $\xi$ are unknown, one possibility to approach the problem is to solve two disjoint equilibrium problems through hard boundary and interface conditions 
\beq
\r^*(r_i)=r_i,\quad
\r^*(\psi)=\xi, \quad
\r(\psi)=\xi, \quad
\r(r_e)=\r_e
\eeq
from which we find distributions of $(\r,\r^*)$ that parametrically depend on $(\psi,\xi)$. To find such parameters for each value of $E$ we thus impose that, at the phase boundary,  
\beq
s_r(\psi) = s_r^*(\psi),\quad s_{\theta}(\psi) = 0. 
\eeq
We now take into consideration two specific examples. 

\begin{figure}[htb]
\begin{centering}
\includegraphics[width=\columnwidth]{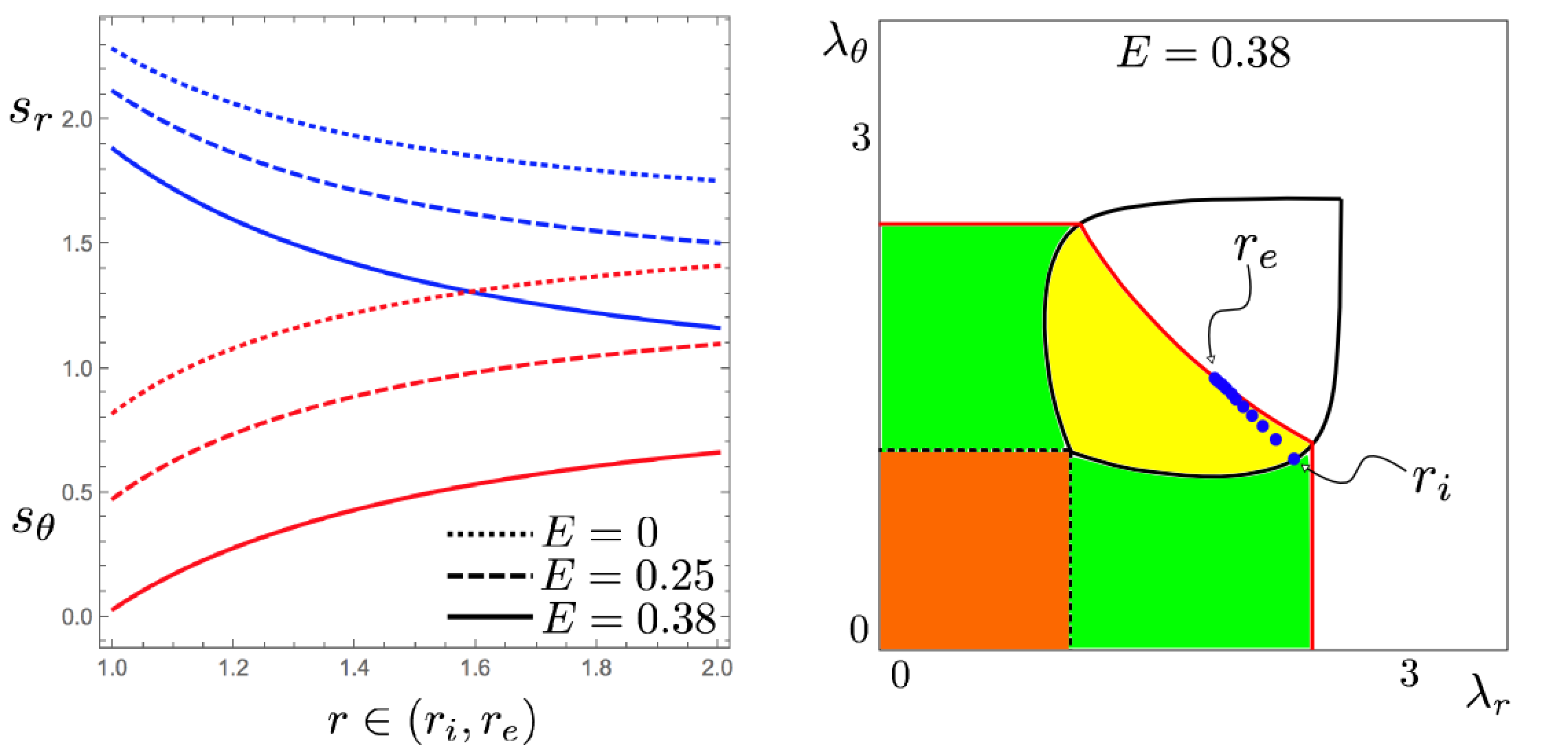}
\par\end{centering}
\caption{\label{case1} Stress distribution (left) and stability plane at failure (right) for a disk constrained as $\r_i=r_i=1$, $r_e=2$ and $\r_e=3$. As the electric field is increased from $E=0$, both stress components relax in the disk. Just immediately prior to the onset of compressive stresses at the internal rim, for $E=0.38$, convexity of the electroelastic energy is lost at the external rim. This is shown in the stability plane (left) where, for $E=0.38$, we report the values of the stretches in the membrane at points $r_k=r_i+k(r_e-r_i)/10$ with $k=1,..,10$. At this critical voltage, convexity is lost at the external rim of the disk.}
\end{figure}

In the first case we set $r_i=1$, $r_e=2$ and $\r_e=3$. Here failure takes place in the taut region through loss of convexity when the electric field reaches the value $E=0.384$. Remarkably, stability is lost when the membrane has not yet started to wrinkle, see Fig.\ref{case1}. 
\begin{figure}[htb]
\begin{centering}
\includegraphics[width=\columnwidth]{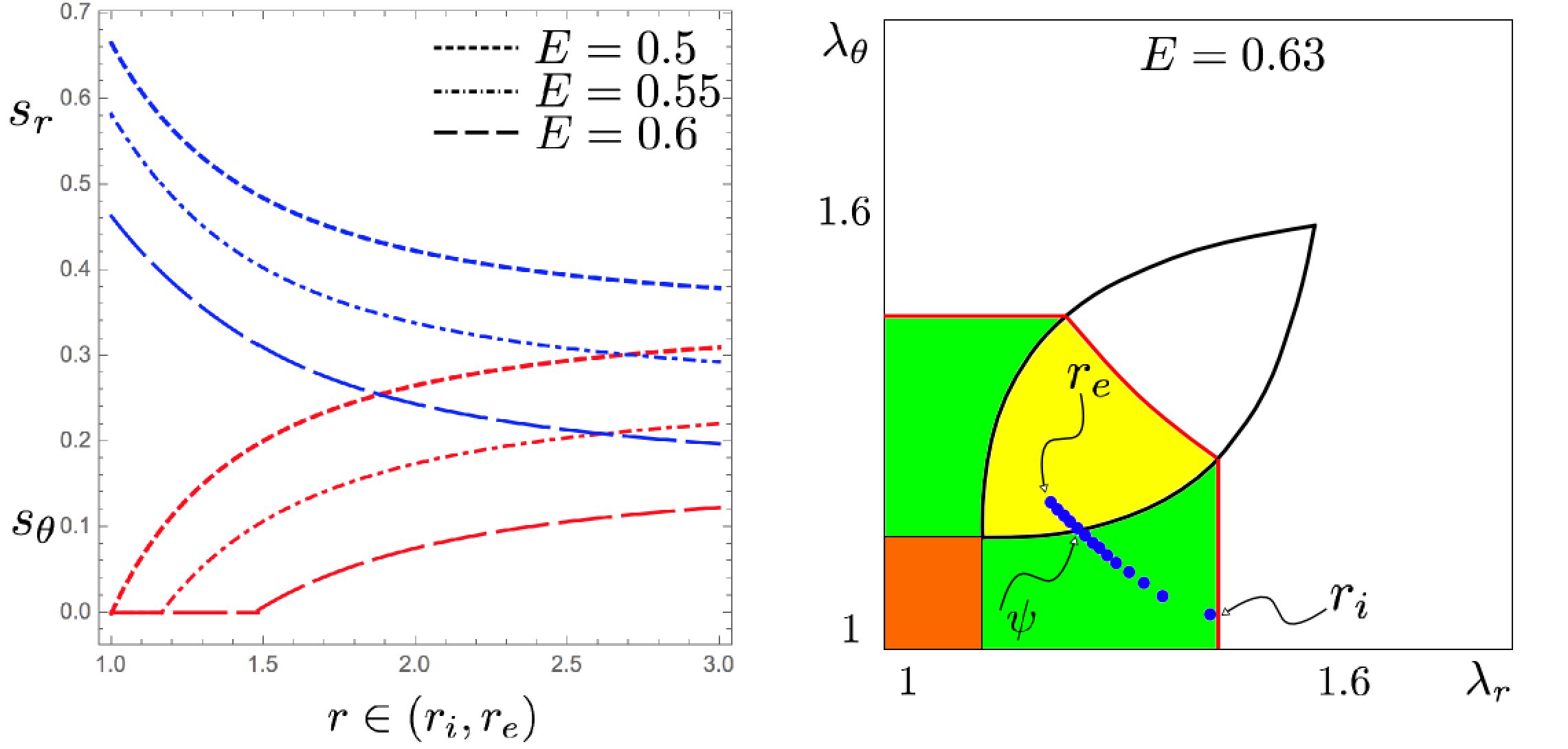}
\par\end{centering}
\caption{\label{case2} Stress distribution (left) and stability plane at failure (right) for a disk with $\r_i=r_i=1$, $r_e=3$ and $\r_e=3.5$. In this case, stress relaxation leads to the appearance of wrinkling at the internal rim for $E=0.5$. The phase boundary then propagates towards the external rim, up until for $E=0.63$ failure occurs due to violation of the tension-extension inequality at the internal rim, see stability plane (left) showing the stretch values at failure at points $r_k=r_i+k(r_e-r_i)/10$ with $k=1,..,10$. Clearly the radius $r=\psi$ is located at the interface between $\mathcal{S}$ and $\mathcal{U}_r$}
\end{figure}

In the second case we set $r_i=1$, $r_e=3$ and $\r_e=3.5$. Now, by increasing the electric field from we again obtain stress relaxation, but convexity of the electroelastic energy still holds when compressive stresses start appearing at the internal rim. As the electric field is further increased, the internal part of the disk wrinkles and the phase boundary between wrinkled and non-wrinkled regions propagates towards the external rim. However, upon reaching the critical electric field $E=0.63$, the tension-extension inequality is violated in the wrinkled part of the disk, at the internal rim. Oppositely to the previous example, in this case stability is lost in the wrinkled region while the taut part of the membrane is still stable, see Fig.\ref{case2}.

\section{Conclusive remarks}

The analysis conducted in this article moves a step towards the modelling of stable wrinkling in voltage and charge controlled dielectric elastomer membranes. Many of our conclusions  are valid under the overly simplified assumption that dielectric elastomers behave as ideal dielectrics, and furthermore that their elastic response is neo-Hookean. In reality few dielectric elastomers have this behavior, above all at high stretches, however the clear understanding of their behavior remains an open issue to date \cite{ZDL}. It should be underlined that also the real meaning of {\it electromechanical instability} is an open issue to date: in our analysis we have neglected the facts that such instabilities are manifested through strong localisation of deformations, and that by employing more realistic material models other complex behaviors may occur, such as phase transitions between thin and thick states \cite{SuoHuang,ZhaoSuo2007} and snap-through instabilities \cite{Su2018}. To bring clarity in these problems is a fundamental step in order to exploit wrinkling in applications, but we leave this challenging task to future investigations.

\end{document}